\documentclass[namedreferences]{SolarPhysics}
\usepackage[optionalrh]{spr-sola-addons} 
\usepackage{graphicx}        
\usepackage{color}           
\usepackage{url}             

\usepackage{solaheader}	
\newcommand{\solareceiveddate}{29 January 2009} 
\newcommand{\solaaccepteddate}{8 June 2009}
\newcommand{\doi}{009-9399-5}
\renewcommand{\solayear}{2009}

\begin{document}
\begin{article}
\begin{opening}

\title{Lifetimes of High-Degree \textit{p} Modes in the Quiet and Active Sun}

\author{O.~\surname{Burtseva}$^{1}$ 
        F.~\surname{Hill}$^{1}$  
	S.~\surname{Kholikov}$^{1}$} 	

\institute{$^{1}$ National Solar Observatory, 950 North Cherry Ave., Tucson, AZ 85719, USA
		  \email{burtseva@noao.edu}}

\author{D.-Y.~\surname{Chou}$^{2}$}

\institute{$^{2}$ Institute of Astronomy and Department of Physics, Tsing Hua University,
Hsinchu, 30043, Taiwan}

\runningauthor{O. Burtseva \textit{et al.}} 
\runningtitle{Lifetimes of High-Degree $p$-Modes}

\begin{abstract}

We study variations of the lifetimes of high-$\ell$ solar \textit{p} modes in the quiet and
active Sun with the solar activity cycle. The lifetimes in the degree range $\ell$ =
300\,--\,600 and $\nu$ = 2.5\,--\,4.5 mHz were computed from SOHO/MDI data in an area
including active regions and quiet Sun using the time\,--\,distance technique. We applied our
analysis to the data in four different phases of solar activity: 1996 (at minimum), 1998
(rising phase), 2000 (at maximum), and 2003 (declining phase). The results from the area with
active regions show that the lifetime decreases as activity increases. The maximal lifetime
variations are between solar minimum in 1996 and maximum in 2000; the relative variation
averaged over all $\ell$ values and frequencies is a decrease of about 13$\%$. The lifetime
reductions relative to 1996 are about 7$\%$ in 1998 and about 10$\%$ in 2003. The lifetime
computed in the quiet region still decreases with solar activity although the decrease is
smaller. On average, relative to 1996, the lifetime decrease is about 4$\%$ in 1998, 10$\%$ in
2000, and 8$\%$ in 2003. Thus, measured lifetime increases when regions of high magnetic
activity are avoided. Moreover, the lifetime computed in quiet regions also shows variations
with the activity cycle.

\end{abstract} 


\end{opening}

\section{Introduction}

The excitation of solar acoustic oscillations is attributed to turbulent convection beneath
the solar surface. The structural and dynamical properties of the Sun, as well as the
processes of excitation and damping of the acoustic modes, change with the solar cycle.
Variation in the excitation and damping are observed as changes in the mode amplitudes and
lifetimes. The study of variations in these parameters provides information on the
mechanism driving the solar oscillations. 

The \textit{p}-mode lifetime in global helioseismology is computed from the line width as
$(2\pi\gamma)^{-1}$, where $\gamma$ is the FWHM of the \textit{p}-mode line profile in
power spectra. Several authors \cite{Chaplin00,JimRey03,Howe03} have shown significant
temporal variations in the mode widths using low-$\ell$ data. They reported a 23$\%$
increase in the mode widths ({\textit i.e.} decrease in the mode lifetimes) with increasing
solar activity. They also found that the variations are frequency dependent, being largest
near 3.1 mHz, and independent of $\ell$. The results have been confirmed with
intermediate-$\ell$ data, showing a 14$\%$ increase in the widths with activity
\cite{Komm00,Salabert06}. 

The authors also analyzed changes in the mode amplitudes, power, and energy-supply rate from
solar minimum to maximum and concluded that the variations probably arise from an increase of
the damping only, since the net forcing of the modes remains constant. Some of the authors
noted that, even if no significant changes in global \textit{p}-mode excitation are evident,
it may possibly change locally. Possible mechanisms, in addition to a continuous turbulent
source of excitation, are transient phenomena such as flares
\cite{Haber88,Kosovichev98,Donea99,Ambastha03,Martinez08,Karoff08} and CMEs
\cite{Gavryusev99}. 

One potential source of damping of the acoustic waves is magnetic activity that is present on
the solar surface. Results of global \cite{Komm02} and local \cite{Rajaguru01, Howe04}
analyses have shown a strong dependence of the mode amplitudes and lifetimes on the locations
of active regions, with amplitude and lifetime decreasing in the presence of strong magnetic
fields. Also, in contrast to the global analysis, Howe \textit{et al.} (2004) reported a
significant decrease of the energy supply rate of \textit{p} modes in active regions. It is
known that the power of the oscillations in sunspots and plages is two\,--\,three times lower
than in the quiet Sun \cite{Woods81}. Possible mechanisms that might be responsible are the
suppression of excitation in sunspots as a strong magnetic field inhibits convection
\cite{Parchevsky07}, absorption of the \textit{p}-mode waves in sunspots
\cite{Braun87,Chen96,Chou96}, the different height of spectral-line formation due to the
Wilson depression, and the modification of \textit{p}-mode eigenfunctions by the magnetic
field \cite{Hindman97}.    

Another possible contribution to \textit{p}-mode damping is variations in the convective
properties near the solar surface. Most likely, these variations arise from the influence of
magnetic structures, as thermal perturbations in the convection zone cannot alone explain
observed changes in irradiance and \textit{p}-mode frequencies over the activity cycle
\cite{Balm96}. Houdek \textit{et al.} (2001) made an attempt to connect changes in the
horizontal length scale of convective eddies and the \textit{p}-mode damping rates and found
that decreasing the horizontal size of convective eddies increases the \textit{p}-mode
damping.

Whatever mechanisms generate and damp solar \textit{p} modes and drive the solar cycle, they
are believed to operate in the convection zone. High-degree ($\ell>200$) acoustic waves
propagate in the upper convection zone and yield information about inhomogeneities in
structure and dynamics of this region. Determination of the line widths and other parameters
of the high-$\ell$ \textit{p} modes has been a problem because of the blending of different
modes in power spectra. Burtseva \textit{et al.} (2007) compared the lifetime of high-$\ell$
\textit{p} modes at solar minimum and maximum using an alternative method based on
time\,--\,distance analysis first proposed by Chou \textit{et al.} (2001). The conclusion of
Burtseva \textit{et al.} (2007) is consistent with low- and intermediate-$\ell$ analysis
(\textit{i.e.}, the lifetime decreases with increasing solar activity). In this work we use
the same method to measure lifetimes of high-degree solar \textit{p} modes in the range
$\ell=300-600$ and $\nu=2.5-4.5$ mHz in four different phases of solar cycle in an area
including active regions and a quiet-Sun area and discuss the results.

\section{Method of Lifetime Measurement} 

The lifetime measured with time\,--\,distance analysis is the lifetime of a wave packet, not
of an individual mode. The wave-packet lifetime is determined by changes in the amplitude and
width of the cross-correlation function of the packet with time. The amplitude of the
cross-correlation function decreases exponentially with the number of skips of the wave below
the photosphere. This effect has been interpreted as the dissipation of solar \textit{p}-mode
power \cite{Chou01}. As a wave packet consists of many \textit{p} modes over some range of
phase velocities ($2\pi\nu/\ell$) dispersion causes it to extend (\textit{i.e.}, its amplitude
decreases and its width increases). In previous studies \cite{Chou07,Burtseva07}, it was found
that the width of the cross-correlation function increases with the number of skips,
demonstrating the effect of dispersion of the wave packet. Dispersion does not change the
energy of the wave packet, which is the product of amplitude squared and width of the wave
packet \cite{Jackson75}. Taking into account both the dissipation and the dispersion, we can
define the lifetime as the e-folding time of the energy of the wave packet due to dissipation
\cite{Chou07,Burtseva07}: 

\begin{equation} 
A_n^2 W_n = A_0^2 W_0 e^{-n\tau_{en}/T} \;, 
\end{equation} 
where $A_n$ and $W_n$ are the amplitude and the width of the cross-correlation of skip $n$,
respectively, $\tau_\mathrm{en}$ is the one-skip envelope time, and $T$ is the lifetime of the wave
packet.

\section{Data Analysis}  

In this study we analyzed two 512-minute series of MDI Dopplergrams in each of the
following epoch: at solar minimum (24 May and 18 September 1996), in the rising phase of
activity (5 and 6 February 1998), at solar maximum (9 and 11 September 2000), and in the
declining phase of activity (12 and 13 December 2003). 

We obtained information on the distribution of magnetic features and the magnetic-field
strength on the analyzed days using MDI 96-minute magnetograms and acoustic-power maps. The
areas including active regions and the quiet areas analyzed in this work are shown in
Figure 1. Concentrations of active regions are denoted with black contours, which
correspond to the magnetic field level of 25 G after smoothing of the images. It is seen
that there is no concentration of active regions in the quiet areas. On both days in 1996
there were no active regions present on the solar disk. On the days in 2000 the Sun was
active with extended active regions, with a relatively low maximal value of magnetic field
of 800 G. On the days in 1998 and 2003 the central part of the Sun was quiet, but a few
small sunspots with magnetic field up to 2000 G were present on the disk at about $40^o$
from the disk center.            

%
\begin{figure}
\centering
\includegraphics[width=0.95\linewidth]{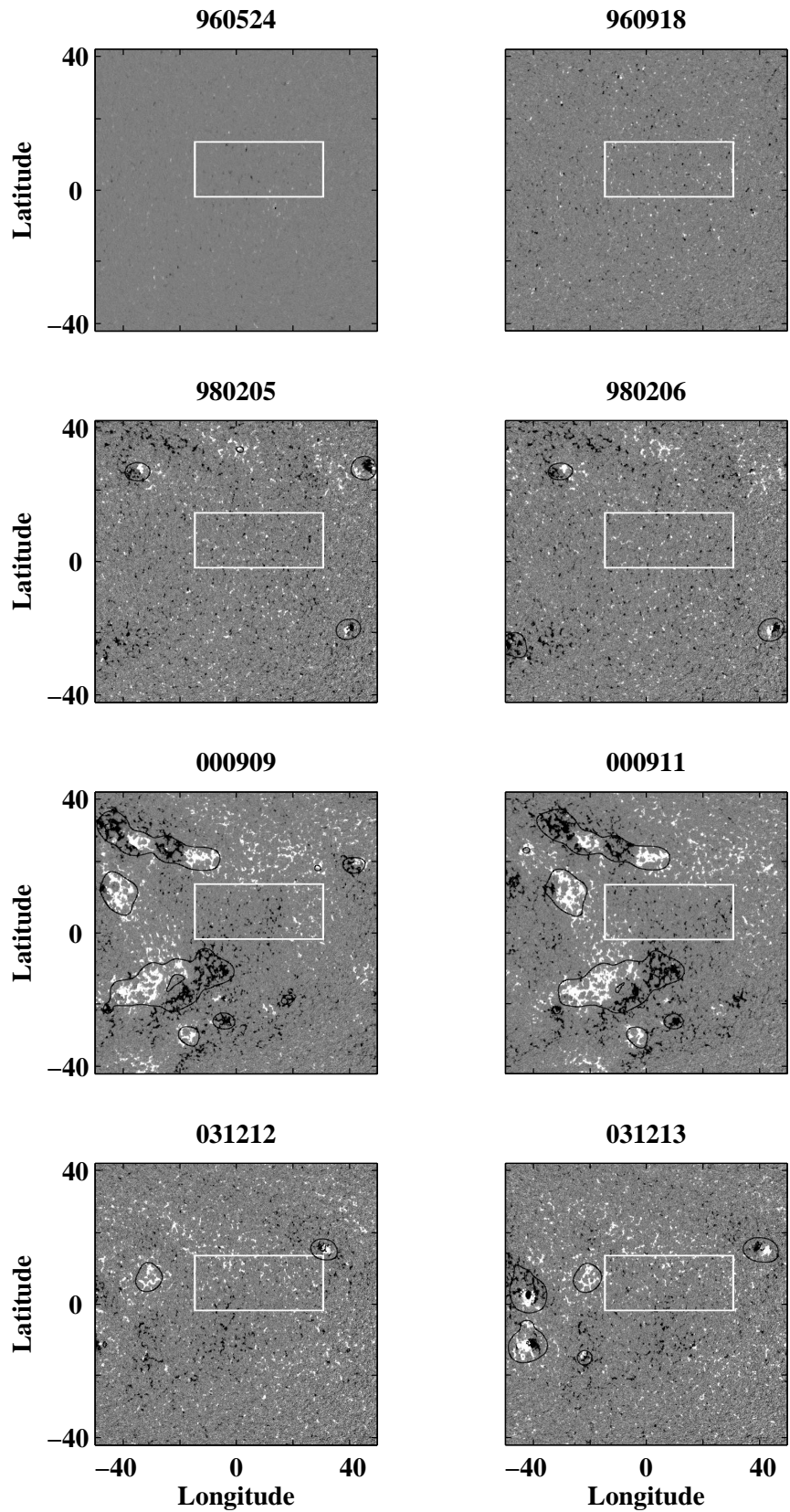}
\caption{MDI magnetograms of the analyzed days. Lifetimes were computed  using  the 
whole  area ($99.4^o \times 85.2^o$), including active regions, and in the quiet area
($45.9^o \times 15.7^o$) shown as a white box around the disk center. Black contours
denote active-region concentrations.} 
\end{figure}

We characterize the activity level on the analyzed days by computing a magnetic activity index
(MAI). The 1$\sigma$ noise level in a single MDI magnetogram is about 20 G \cite{Scherrer95}.
To reliably compute MAI we first set all unsigned flux less than 2.5$\sigma = 50$ G to zero
and correct images for outliers using a technique similar to that described in \cite{Basu04}.
We compute MAI by summing the absolute values of the magnetic flux inside the analyzed areas
and normalizing the sum by the total number of pixels in the areas instead of the number of
nonzero pixels. We also compute MAI using NSO/Kitt Peak magnetograms (KPM) for comparison. The
1$\sigma$ noise level of KPM data is about 5 G \cite{Wenzler04}. Similarly, we set all
unsigned flux less than 2.5$\sigma = 12$ G to zero. Then we multiply the KPM images by 1.7 to
correct for the difference in scale as MDI shows a larger magnetic flux than KPM. The computed
MAIs are shown in Table 1. The MDI magnetograms for 24 May 1996 and KPM data for the two days
in 2003 are absent.   

The data analysis and cross-correlation function computation procedure is similar to that
used in \cite{Burtseva07} and briefly described as follows:

\begin{table}
\caption{Unsigned magnetic field (in Gauss) averaged over the whole area including active regions and the quiet area (see Figure 1). }
\begin{tabular}{c@{ } c@{ }c r@{ $\pm$ }l r@{ $\pm$ }l r@{ $\pm$ }l r@{ $\pm$ }l}
\hline
\multicolumn{3}{c}{Date} &\multicolumn{2}{c}{Area including} &\multicolumn{2}{c}{Quiet Area} &\multicolumn{2}{c}{Area including} &\multicolumn{2}{c}{Quiet Area}\\
\multicolumn{3}{c}{}     &\multicolumn{2}{c}{active regions} &\multicolumn{2}{c}{(MDI)}	     &\multicolumn{2}{c}{active regions} & \multicolumn{2}{c}{(KPM)}\\
\multicolumn{3}{c}{}     &\multicolumn{2}{c}{(MDI)}	     &\multicolumn{2}{c}{ }	     &\multicolumn{2}{c}{(KPM)} 	 & \multicolumn{2}{c}{} \\
\hline
24&May&1996 &\multicolumn{2}{c}{\hspace{-1.3mm}\,--\,} &\multicolumn{2}{c}{\,--\,} &1.73 &0.03 &2.49 &0.11 \\
18&Sep&1996 &2.43  &0.02 &1.90 &0.07 &1.98  &0.03 &2.80 &0.11 \\
05&Feb&1998 &3.60  &0.04 &1.71 &0.04 &3.25  &0.07 &2.15 &0.11 \\
06&Feb&1998 &4.45  &0.06 &2.02 &0.07 &3.01  &0.08 &2.05 &0.13 \\
09&Sep&2000 &15.03 &0.33 &5.40 &0.12 &15.00 &0.12 &6.75 &0.20 \\
11&Sep&2000 &16.21 &0.33 &4.93 &0.17 &15.20 &0.11 &5.42 &0.16 \\
12&Dec&2003 &3.99  &0.26 &3.37 &0.14 &\multicolumn{2}{c}{\hspace{-1.3mm}\,--\,} &\multicolumn{2}{c}{\,--\,} \\
13&Dec&2003 &5.85  &0.27 &3.53 &0.14 &\multicolumn{2}{c}{\hspace{-1.3mm}\,--\,}	&\multicolumn{2}{c}{\,--\,} \\
\hline
\end{tabular}
\end{table}

\begin{itemize}

\item[\textit{i}\textrm{)}] Images are remapped, tracked, filtered with 15-minute running
mean, and transformed into the ($\ell$, $m$, $\nu$) domain, where $m$ is the azimuthal
degree. 

\item[\textit{ii}\textrm{)}] A Gaussian filter of FWHM = 2.0 mHz centered at a frequency
$\nu_0$ is applied. 

\item[\textit{iii}\textrm{)}] A phase-velocity filter is applied to isolate the modes in a
range of angular phase velocity ($w=2\pi\nu/\ell$) forming a wave packet with the central
frequency $\nu_0$ and the corresponding degree $\ell_0$. The filter is smoothed with a
Hanning window in frequency.

\item[\textit{iv}\textrm{)}] The data are reconstructed back to the space-time domain.

\item[\textit{v}\textrm{)}] The center-annuli cross-correlation $C(\Delta,\tau)$ is
computed with 

\begin{equation}                                                               
C(\Delta,\tau)=\int\Psi(0,t)\Psi(\Delta,t+\tau)\mathrm{dt},                             
\end{equation}                           

where $\Psi(0,t)$ is the signal measured at the central point at time $t$ and
$\Psi(\Delta,t+\tau)$ is the signal averaged over an annulus at an angular distance
$\Delta$ from the central point at time $t+\tau$. The procedure is repeated for different
central points at the disk center. The cross-correlation functions of different central
points are averaged. 

\item[\textit{vi}\textrm{)}] For each skip, $C(\Delta,\tau)$ is fitted by a Gabor wavelet
\cite{Kos&Duv97}

\begin{equation}
G(\tau)=A\cdot \mathrm{cos}[2\pi\nu(\tau-\tau_{ph})] \cdot
\mathrm{exp}\bigg[-\frac{(\tau-\tau_{en})^2}{2W^2}\bigg],
\end{equation}

where $A$, $\nu$, and $W$ are the amplitude, frequency, and width of a Gaussian envelope,
respectively. The parameters $\tau_\mathrm{ph}$ and $\tau_\mathrm{en}$ are the phase time and
envelope time, respectively. For each skip, $A$, $W$, and $\tau_\mathrm{en}$ are averaged over
five angular distances near the maximum of the cross-correlation amplitude to reduce noise.
The procedure is repeated for both positive and negative $\tau$. The values of $A$ and $W$ are
the averages of results for positive and negative $\tau$, since they are found to be very
close in magnitude.

\item[\textit{vii}\textrm{)}] The lifetime ($T$) is determined  by fitting $\mathrm{ln}\, (A^2
W)$ \textit{versus} $n$ according to Equation (1). The details are discussed in Burtseva
\textit{et al.} (2007).   

\end{itemize}

The above analysis was performed for the whole area including active regions and the quiet
area near the disk center shown in Figure 1. A minimum four-skip distance is required to
compute lifetimes with this method. The size of the quiet area does not allow us to perform
the analysis for the wave packets with $\ell_{0}=300$, $\nu_{0}=2.5-4.5$ mHz and
$\ell_{0}=400$, $\nu_{0}=4.0$ and $4.5$ mHz as the distance required for the lifetime
computation for those waves is from about $13^\mathrm{o}$ to $24^\mathrm{o}$. For the quiet area we
have about $7^\mathrm{o}$ maximum, thus we are able to compute lifetimes only for the wave packets
$\ell_{0}=400$, $\nu_{0}=2.5-3.5$ mHz $\ell_{0}=500-600$, $\nu_{0}=2.5-4.5$ mHz which
requires distances from about $5^\mathrm{o}$ to $7^\mathrm{o}$.

\section{Results and Discussion}

%
\begin{figure} 
\centering
\includegraphics[width=0.95\linewidth]{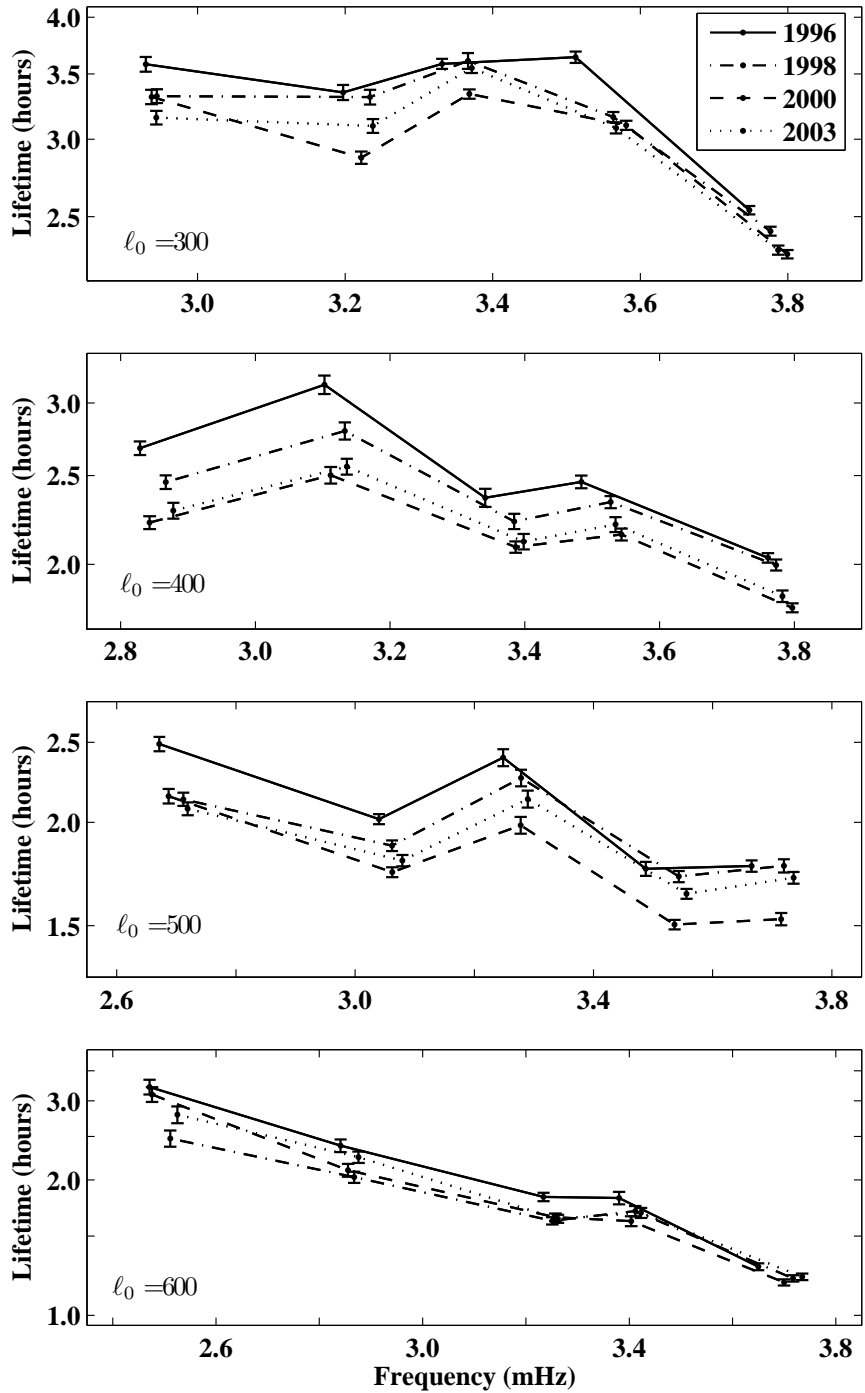}
\caption{\textit{p}-mode lifetime computed in the area including active regions, averaged
over the results from two 512-minute time series in each epoch \textit{versus} frequency
for various $\ell_0$. The frequency is determined from the Gabor fit for the first-skip
cross-correlation function.}
\end{figure}

\begin{figure} 
\centering 
\includegraphics[width=0.95\linewidth]{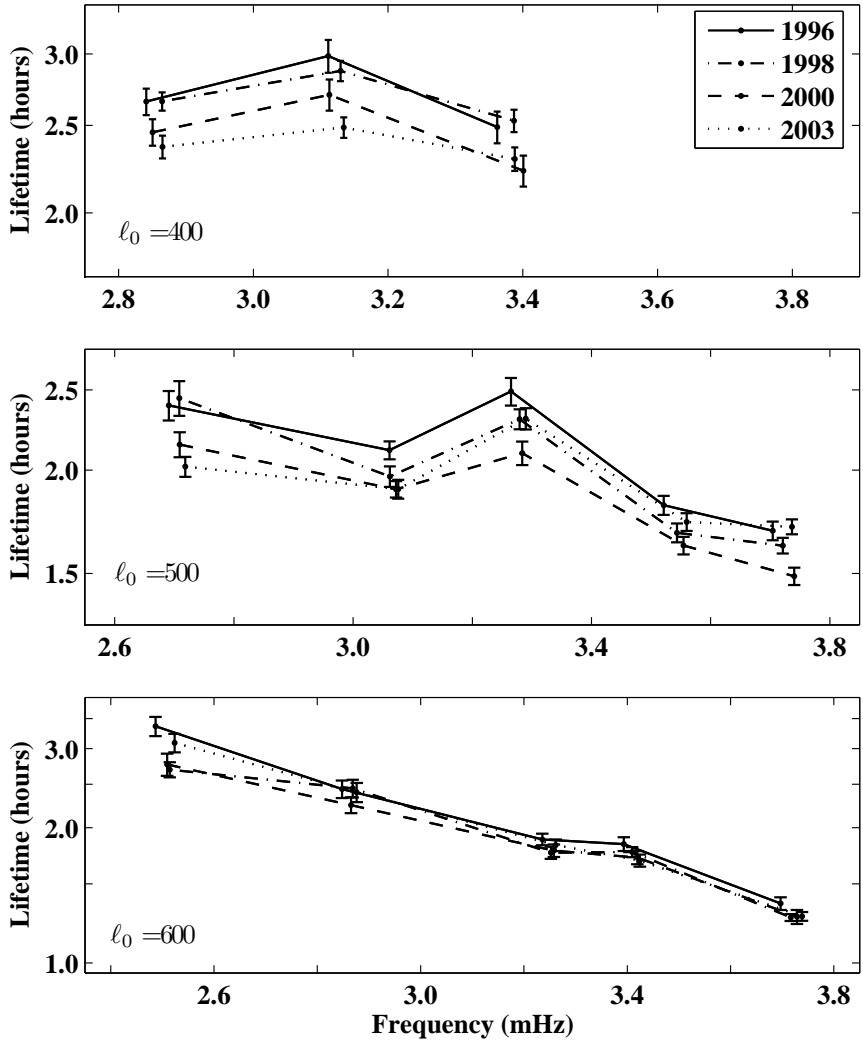}
\caption{\textit{p}-mode lifetime computed in the quiet area, averaged over the results
from two 512-minute time series in each epoch \textit{versus} frequency for various
$\ell_0$. The frequency is determined from the Gabor fit for the first-skip
cross-correlation function.}
\end{figure}  
%

\begin{table} [!hb] 

\caption{Percentage of the lifetime decrease relative to 1996, averaged over
all $\ell$ values and frequencies of 3.0, 3.5, and 4.0 mHz.}

\label{T-simple}
\begin{tabular}{ccc}     
  \hline                   
Year/1996 & Area including active regions ($\%$) & Quiet Area ($\%$) \\
  \hline
1998/1996 & ~6.9 & 4.2 \\
2000/1996 & 13.2 & 9.6 \\
2003/1996 & ~9.6 & 7.6 \\
  \hline
\end{tabular}
\end{table}

The lifetimes of solar \textit{p} modes computed in the area including active
regions are shown in Figure 2. It is seen that the lifetime decreases as activity
increases.The results are in agreement with all previous studies of the
\textit{p}-mode lifetimes mentioned earlier. The percentage of the lifetime decrease
for all analyzed days relative to the days in 1996 is listed in Table 2. The maximal
lifetime variations are between solar minimum in 1996 and maximum in 2000; the
relative variation averaged over all $\ell$ values and frequencies is about 13$\%$. The
lifetimes in 2003 are closer to the lifetimes in 2000, and the lifetimes in 1998 are
closer to those in 1996. The lifetime variations relative to 1996 are about 7$\%$ in
1998 and about 10$\%$ in 2003. As seen in Table 1, the activity level was higher on
the analyzed days in 2003 than in 1998.        

The lifetimes computed in the quiet area are presented in Figure 3. After excluding active
regions from the analysis, the lifetimes decrease with solar activity, although the
decrease is smaller. On average, relative to 1996, the decrease in the lifetime is about
4$\%$ in 1998, 10$\%$ in 2000, and 8$\%$ in 2003. The daily variations of the lifetime in
1996 are within the uncertainties of the measurements. 

Thus, excluding high magnetic activity regions increases the measured lifetime. From the
difference between lifetimes and the magnetic field strength of the quiet area and the area
including active regions, and also from the distribution of the active features on the
solar disk on the analyzed days (see Figure 1), we speculate that acoustic waves are
affected more by large active regions as seen at solar maximum, and less by small sunspots
with strong magnetic field. This is likely because the wavelengths of the acoustic waves
are typically larger than the smaller sunspots and better matched with the large active
regions. 

The lifetime computed in the quiet area still shows variations with activity cycle. What could
cause the variations in the quiet area where there are no large-scale magnetic field
concentrations? One explanation could be that the convective properties change with solar
cycle. As discussed in the Introduction, Houdek \textit{et al.} (2001) found that damping of
\textit{p} modes increases as horizontal size of convective eddies decreases. It has been
reported by some authors \cite{Muller88,Berrilli99} that the horizontal scale of the solar
granules decreases from solar minimum to maximum, which perhaps indicates that the horizontal
size of the convective eddies does indeed decrease. Meunier \textit{et al.} (2008) reported
that supergranular cells are smaller at solar maximum. However, other studies arrive at the
contradictory conclusion that granular size increases with activity
\cite{Hanslmeier02,Saldana04}.  

Solar-cycle variation of the lifetime in the quiet area caused by purely structural
perturbations in the convection zone should result in lifetime variations with no
correlation with magnetic field strength. To test this we computed the average lifetime for
each wave packet over the four epochs and normalized each of the lifetimes by the average.
Then we plot the normalized lifetimes of all wave packets analyzed \textit{versus} MAI
computed in the area including active regions and the quiet area in Figure 4. The lifetimes
obviously anticorrelate with MAI for the area including active regions. In the quiet area
we also see an inverse correlation of the lifetimes with MAI. This suggests that the
magnetic field somehow plays a role in the lifetime variation of \textit{p} modes in the
quiet Sun. Future work will involve more statistics to confirm the
conclusions.               

\begin{figure} 
\centering 
\includegraphics[width=0.95\linewidth]{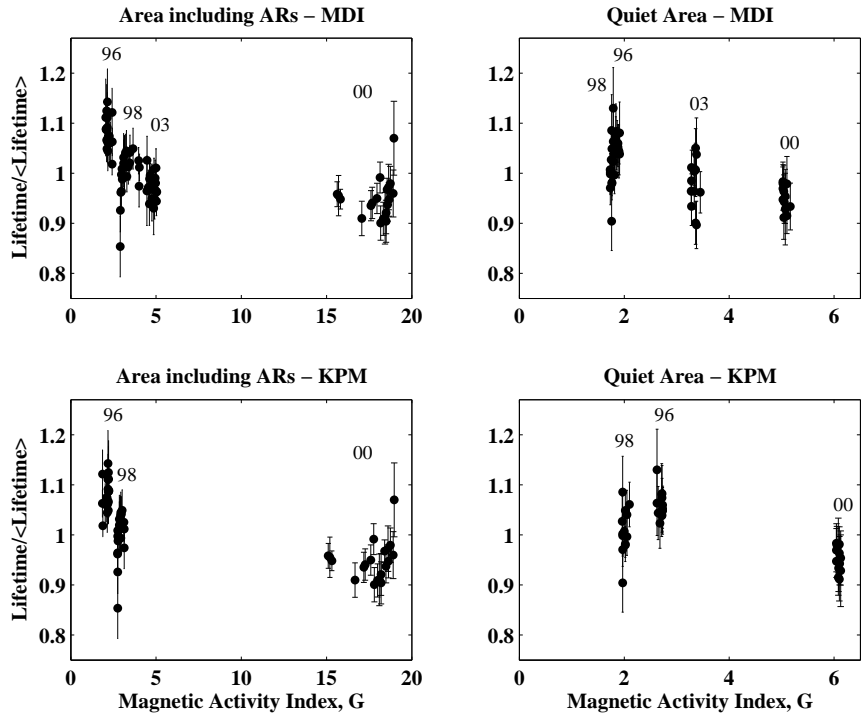}
\caption{Normalized lifetimes \textit{versus} magnetic activity index in the area including
active regions (left) and the quiet area (right). Top and bottom panels correspond to the
MAI computed from MDI and Kitt Peak magnetograms, respectively.}  
\end{figure}  

All of the analyzed quiet regions have different levels of magnetic field (see Table 1).
Some isolated pixels in the quiet regions have magnetic fields as strong as $\approx$500
G, which are the strong-field component of the network field at the supergranular
boundaries \cite{Harvey71,Lin95}.  

Thus, the elements of heightened magnetic activity in the quiet regions could cause the
residual lifetime variations in the quiet area. The MAIs of the analyzed quiet areas also show
solar-cycle variations. One can clearly see the increase in MAI from Table 1 (plotted in
Figure 5) corresponding to the days at solar maximum for both the area with active regions and
the quiet area. Most authors who analyzed solar-cycle variations of the irradiance and
photospheric magnetic field reported little overall solar-cycle variations in the quiet-Sun
network elements arising from decaying active regions, and no variations in the intranetwork
field \cite{Harvey94,Topka97,Pevtsov01,Hagenaar03,Sanchez03,Pauluhn03}. Muller \textit{et al.}
(2007) reported solar-cycle related variations of the granulation contrast at the solar
surface connected to the temperature differences between granules and intergranular lanes. The
contrast variations may be due to changes in the physical structure under the surface or a
varying amount of magnetic flux in the intergranular lanes.      

The recently discovered horizontal magnetic field in quiet regions in the solar
photosphere \cite{Harvey07,Lites08} might also affect the propagation of the acoustic
waves in the quiet Sun in a way different from that of the vertical component of the
magnetic field. As the authors speculate, the possible driving mechanism for the
horizontal field is granular and supergranular convection and transformation of the field
lines in response to the evolving flux distribution in the network and intranetwork field
in the quiet Sun.

In future work, we will analyze a large number of quiet areas to confirm temporal
variations of the lifetimes of high-$\ell$ acoustic waves found in this work. We will also
study variability of the apparent power of the acoustic modes in an area including active
regions and a quiet area in different phases of the solar cycle.  

\begin{figure} 
\centering 
\includegraphics[width=0.95\linewidth]{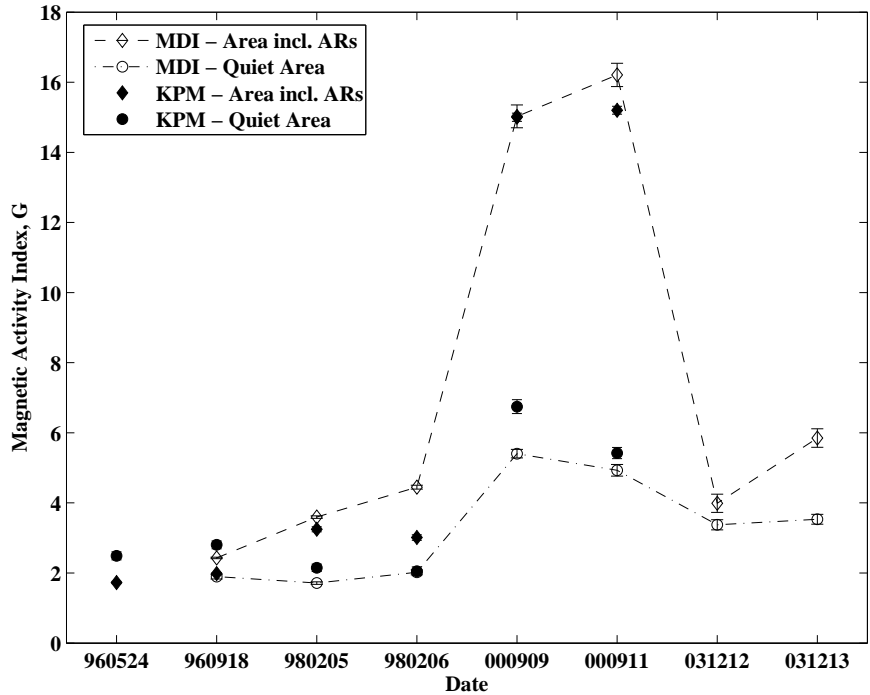}
\caption{Temporal evolution of MAIs from Table 1 vs. four epochs used in the lifetime
study. The line and symbol styles are shown in the top-left corner of the figure. The plot
clearly shows solar cycle variations of the MAI in both the area including active regions
and the quiet area.}  
\end{figure}  

\section*{Acknowledgments}

The National Solar Observatory is operated by AURA, Inc. under a cooperative agreement with
the National Science Foundation. SOHO is a mission of international cooperation between ESA
and NASA. NSO/Kitt Peak data used here are produced cooperatively by NSF/NSO, NASA/GSFC,
and NOAA/SEL. The authors thank R. Bogart, J. Harvey, C. Lindsey, J.C. Martinez-Oliveros, 
A. Norton, G. Petrie, N.-E. Raouafi, and S. Tripathy for fruitful discussions of the results
of this work, helpful suggestions, and comments.

\end{article}
\end{document}